\documentclass{desyproc}
\begin{document}
\title{Prospects of Search for Solar Axions with Mass over $1 \ \mathrm{eV}$ and Hidden Sector Photons}

\author{{\slshape R. Ohta$^{1,6}
$, M. Minowa$^{1,5}$, Y. Inoue$^{2,5}$, Y. Akimoto$^3$, T. Mizumoto$^1$,
A. Yamamoto$^{4,5}$}\\[1ex]
$^1$Department of Physics, School of Science, the University of Tokyo\\
$^2$International Center for Elementary Particle Physics, theUniversity of Tokyo\\
$^3$Graduate School of Medicine and Faculty of Medicine, the University of Tokyo\\
$^4$High Energy Accelerator Research Organization (KEK),
  1-1~Oho, Tsukuba, Ibaraki 305-0801, Japan\\
$^5$Research Center for the Early Universe (RESCEU), School of Science,
the University of Tokyo\\
$^6$Research Fellow of the Japan Society for the Promotion of Science\\
$^{1,2,3,4}$ 7-3-1 Hongo, Bunkyo-ku, Tokyo 113-0033, Japan}

\contribID{ryosuke\_ohta}

\desyproc{DESY-PROC-2009-05}
\acronym{Patras 2009} 
\doi  

\maketitle

\begin{abstract}
We present prospects of two experiments using the Tokyo Axion Helioscope.
One is a search for solar axions.
In the past measurements, axion mass from 0 to $0.27 \ \mathrm{eV}$ and from 0.84 to $1.00 \ \mathrm{eV}$ have been scanned and no positive evidence was seen. 
We are now actively preparing a new phase of the experiment aiming at
 axion mass over $1 \ \mathrm{eV}$. 
The other is a search for hidden sector photons from the Sun. 
We have been designing and testing some additional equipments, which have to be installed on the helioscope to search for hidden photons with mass of over $10^{-3} \ \mathrm{eV}.$
\end{abstract}

\section{Introduction}
The Sun could copiously emit weakly interacting particles, that could
eventually be detected inside a sensitive detector at the Earth.

\begin{wrapfigure}{r}{6cm}
  \includegraphics[scale=0.09]{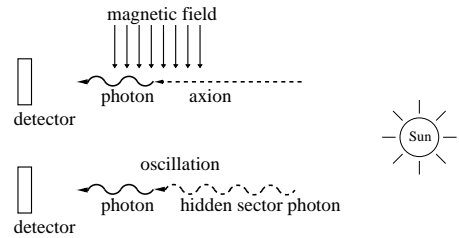}
  \caption{Detection schematics of solar axions and hidden photons from the Sun.}
 \label{Fig:incident}
\end{wrapfigure}

Axion is one of such particles.
The existence of axion is implied to solve the strong CP problem
~\cite{Peccei:1977hh}. 
Axions are expected to be produced in the solar core through their coupling to photons.
This process is called Primakoff process. 
The outgoing axion has average energy of about 4 keV~\cite{Raffelt:2005mt}.
Sikivie proposed an ingenious experiment to detect such axions~\cite{Sikivie:1983ip}. 
A detection schematics of solar axion is shown in Fig.~\ref{Fig:incident}. 
The detection device called axion helioscope is a system of a strong magnet and an X-ray detector, where the solar axions are transformed into X-ray photons through the inverse Primakoff process in the magnetic field. 
Conversion is coherently enhanced even for massive axions by filling the conversion region with light gas. 
If the axion mass $m_a$ is at around a few eV, detection of the solar axion becomes feasible.

Hidden sector photon is another kind of weakly interacting particles. 
The existence of hidden photons is predicted by several extensions of Standard Model. 
If light hidden photons exist, they could be produced through kinetic
mixing with solar photons~\cite{Redondo:2008aa, Gninenko:2008pz}.
Therefore it is natural to consider the Sun as a source of low energy hidden photons. 
A detection schematics of hidden photon from the Sun is also shown in Fig.~\ref{Fig:incident}. 
Unlike the case of the axion, no magnetic field is required to transform photons into hidden photons. 

In this paper we report the current status of two experiments.
One is the search for solar axions and the other is the search for hidden photons from the Sun.

\section{Tokyo Axion Helioscope}
\begin{figure}[t]
  \begin{minipage}[t]{0.5\textwidth}
    \hrule height 0pt
    \includegraphics[scale=0.8]{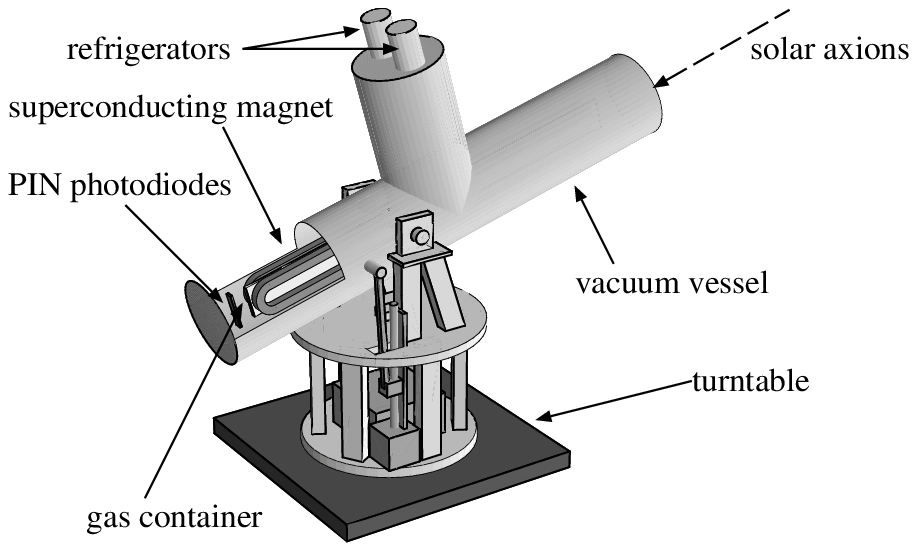}
    \caption{The schematic view of the axion helioscope.}
    \label{Fig:sumico}
  \end{minipage}
  \hskip 0.02\textwidth 
  \begin{minipage}[t]{0.4\textwidth}
    \hrule height 0pt
    \includegraphics[scale=0.7]{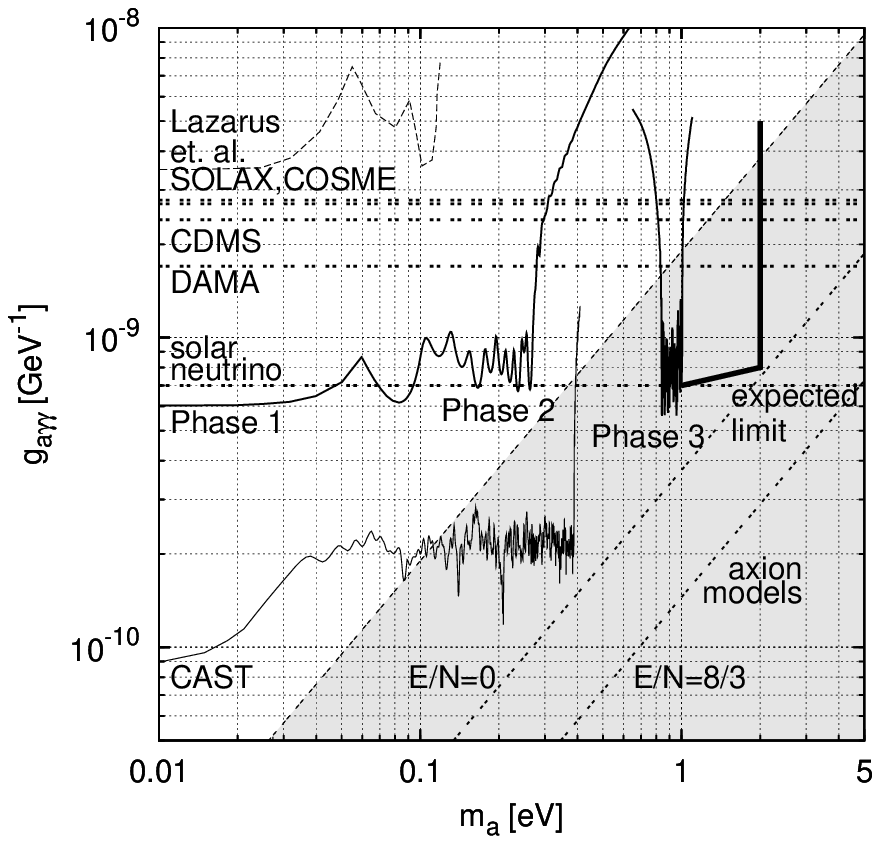}
    \vspace{-10mm}
    \caption{Exclusion limit on $g_{a\gamma\gamma}$ versus $m_a$
      at 95\% confidence level.
    }
    \label{Fig:sumicoresult}
  \end{minipage}
\end{figure}

The schematic figure of the axion helioscope is illustrated in Fig.~\ref{Fig:sumico}. 
It consists of a superconducting magnet, X-ray detectors, a gas container, and an altazimuth mounting. 
The magnet~\cite{Sato:1997} consists of two 2.3-m long race-track shaped superconducting coils running parallel with a 20-mm wide gap between them. 
The transverse magnetic field in the gap is $4 \ \mathrm{T}$. 
The magnetic field can be maintained without an external power supply with a help of a persistent current switch. 
The magnet is kept lower than $6 \ \mathrm{K}$ by two Gifford-McMahon refrigerators. 
The container to hold dispersion-matching gas is inserted in the aperture of the magnet. 
It is made of four 2.3-m long stainless-steel square pipes and 5N high purity aluminium sheets wrapping around them to achieve high uniformity of temperature. 
The measured thermal conductance between the both ends was $1 \times 10^{-2} \ \mathrm{W/K}$ at $6 \ \mathrm{K}$ under $4 \ \mathrm{T}$. The one end of the gas container is suspended by three Kevlar cords. 
The other end at the opposite side is flanged to the magnet. This end is terminated with an X-ray window which is transparent above $2 \ \mathrm{keV}$ and can hold gas up to $0.3 \ \mathrm{MPa}$. 
The gas introducing pipelines are also at this side and have an automated gas controlling system which enables us to scan wide range of axion mass. 
The generated X-ray is viewed by sixteen PIN photodiodes. 
Details on the X-ray detector are given in Ref.~\cite{Namba:2001fz, Akimoto:2005rf}.
Except for the gas controlling system, they are constructed in a vacuum vessel which is mounted on an altazimuth mount to track the Sun. 
It can track the Sun about a half of a day.
During the other half of a day, background spectrum is measured. 

Phase 1 of the solar observation was performed in December 1997 without  the gas container~\cite{Moriyama:1998kd}. 
Phase 2 was performed from July to September 2000 with the gas container and low density helium gas~\cite{Inoue:2002qy}.
Phase 3 was performed from December 2007 to April 2008 with higher density helium gas than that of Phase 2~\cite{Inoue:2008zp}.
Since those measurements result in no positive signals of axion, upper limits on the axion-photon coupling constant $g_{a\gamma\gamma}$ were set to be $g_{a\gamma\gamma} < 6.0 - 10.4 \times 10^{-10} \ \mathrm{GeV}^{-1}$ for $m_a < 0.27 \ \mathrm{eV}$ and $g_{a\gamma\gamma} < 5.6 - 13.4 \times 10^{-10} \ \mathrm{GeV}^{-1}$ for $0.84 < m_a < 1.00 \ \mathrm{eV}$.
We are now preparing the search for solar axion with mass over 1 eV introducing higher density helium gas than that of last phase. 
Figure~\ref{Fig:sumicoresult} shows the expected upper limit of next measurement. Our previous limits and the some other bounds are also plotted in the same figure. 
The SOLAX~\cite{Gattone:1997hf}, COSME~\cite{Bernabei:2001ny}, DAMA~\cite{Morales:2001we} and CDMS~\cite{Collaboration:2009ht} are solar axions experiments which exploit the coherent conversion on the crystalline plains in germanium and a NaI detector.
The experiment by Lazarus et al.~\cite{Lazarus:1992ry} and CAST
~\cite{Zioutas:2004hi, Andriamonje:2007ew, Arik:2008mq} are the same kind of experiments as ours.
The limits $g_{a\gamma\gamma} < 7 \times 10^{-10} \ \mathrm{GeV}^{-1}$ is the solar limit inferred from the solar neutrino flux consideration~\cite{Gondolo:2008dd}.
Preferred axion models~\cite{Kaplan:1985dv, Srednicki:1985xd,
Cheng:1995fd} are also shown by the shaded area in the Fig.~\ref{Fig:sumicoresult}.

\section{Search for hidden photon}
 \begin{wrapfigure}{r}{7cm}
  \vspace{-5mm}
  \includegraphics[scale=0.215]{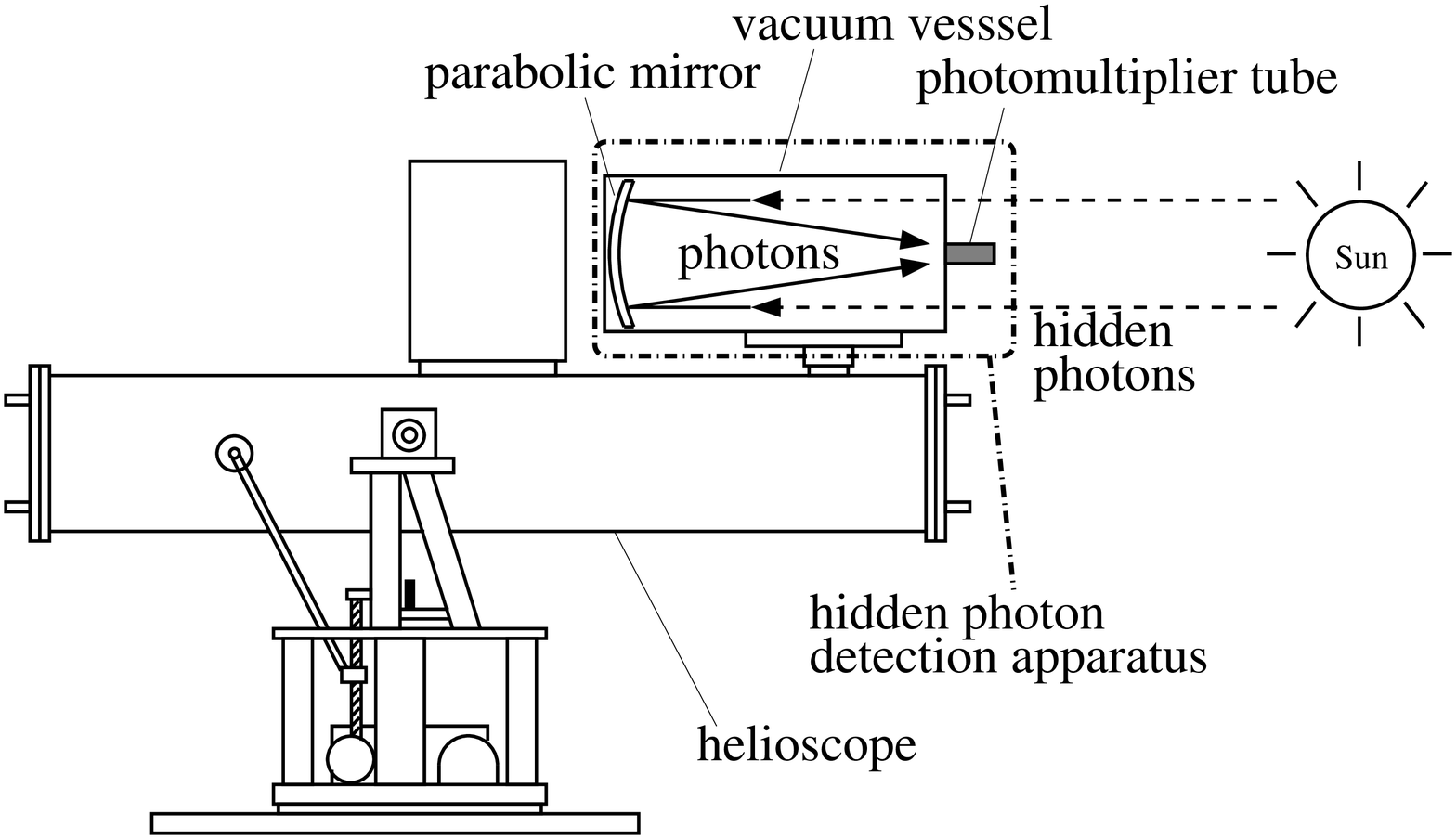}
  \vspace{-3mm}
  \caption{The schematic view of the apparatus to search for hidden photon from the Sun.}
  \label{Fig:hiddenschematic}
  \vspace{5mm}
  \includegraphics[scale=0.235]{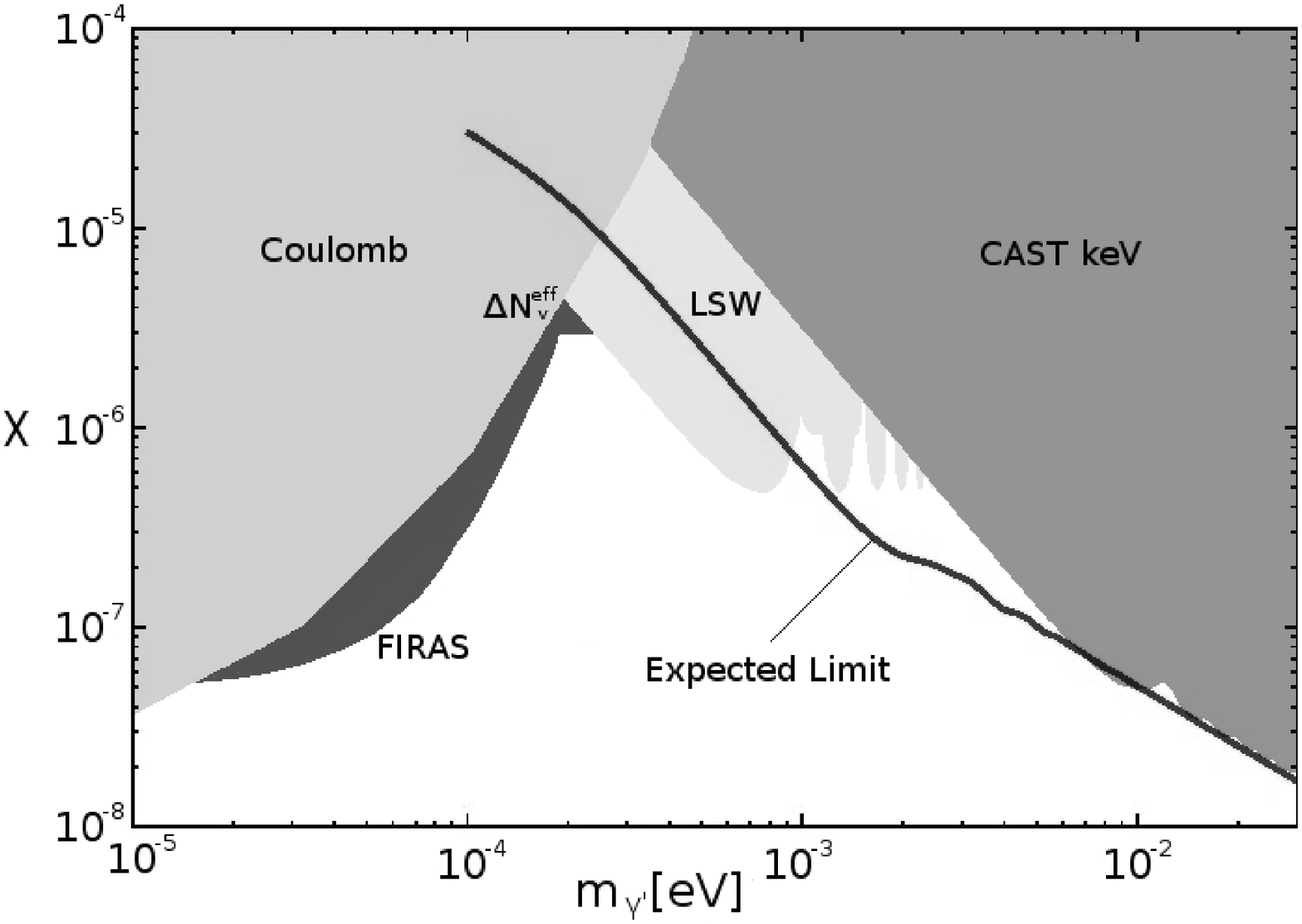}
  \vspace{-5mm}
  \caption{Exclusion limit on mixing strength between photon and hidden
 photon $\chi$ versus hidden photon mass $m_{\gamma^\prime}$. 
The solid line shows our expected limit. }
 \label{Fig:hiddenexpected}
 \end{wrapfigure}

To search for hidden photons from the Sun, we plan to add an additional apparatus on the cylinder of the helioscope. 
A schematic design of the apparatus is illustrated in Fig.~\ref{Fig:hiddenschematic}.
It mainly consists of a vacuum vessel as a conversion region, a parabolic mirror, and a photomultiplier (PMT). 
In one side of the vessel, the parabolic mirror is attached to collect
photons produced from the hidden photon - photon oscillation and the focal point of the mirror is set at the other side of the vessel. 
The mirror has a diameter of 50 cm, and a focal length of 1 m. 
On the focal point, the PMT is attached to detect collected photons. 
In addition, we plan to cool the PMT to reduce dark count rate.
As a preliminary experiment, we have cooled R329-02, a product of Hamamatsu photonics, and measured its dark count rate.
The measured rate at $-30 \ ^o\mathrm{C}$ was about $10 \ \mathrm{Hz}$. 
This rate is several times lower than a dark count rate at room temperature.
For actual experiment, we plan to use a more suitable one than R329-02.

If we suppose the dark count rate is $10 \ \mathrm{Hz}$, pressure in the
vessel is much less than $10 \ \mathrm{Pa}$, the length of conversion
region is $1 \ \mathrm{m}$, the diameter of the mirror is $0.5 \
\mathrm{m}$, reflectivity of the mirror is $90 \ \%$, and measuring time
is $10^6 \ \mathrm{s}$, we expect an exclusion region above the solid line shown in Fig.~\ref{Fig:hiddenexpected}. 
The limits from other experiments and observations: Coulomb's law tests~\cite{Williams:1971ms, Bartlett:1988yy},
``light shining through walls'' experiments~\cite{Ahlers:2007qf, Afanasev:2008fv}, 
CAST~\cite{Redondo:2008aa}
and exclusion from CMB observation~\cite{Jaeckel:2008fi, Redondo:2008zf} are also shown in Fig.~\ref{Fig:hiddenexpected}.

\section*{Acknowledgments}
The authors thank the former director general of KEK, Professor H. Sugawara,
for his support in the beginning of the helioscope experiment.
This research was partially supported
by the Japanese Ministry of Education, Science, Sports and Culture,
Grant-in-Aid for COE Research and Grant-in-Aid for Scientific Research (B),
and also by the Matsuo Foundation.

\begin{footnotesize}

\end{footnotesize}



\begin{thebibliography}{10}

\bibitem{Peccei:1977hh}
R.~D. Peccei and Helen~R. Quinn.
\newblock {\em Phys. Rev. Lett.}, 38:1440--1443, 1977.

\bibitem{Raffelt:2005mt}
Georg~G. Raffelt.
\newblock {arXiv:hep-ph/0504152}.
\newblock 2005.

\bibitem{Sikivie:1983ip}
P.~Sikivie.
\newblock {\em Phys. Rev. Lett.}, 51:1415, 1983.

\bibitem{Redondo:2008aa}
Javier Redondo.
\newblock {\em JCAP}, 0807:008, 2008.

\bibitem{Gninenko:2008pz}
Sergei~N. Gninenko and Javier Redondo.
\newblock {\em Phys. Lett.}, B664:180--184, 2008.

\bibitem{Sato:1997}
Y~Sato et~al.
\newblock {\em Proc. of the 15th International Conference on Magnet Technology
  (MT-15) ed Liangzhen L, Guoliao S and Luguang Y (Beijing: Science Press)},
  pages 262--265, 1998.

\bibitem{Namba:2001fz}
T.~Namba, Y.~Inoue, S.~Moriyama, and M.~Minowa.
\newblock {\em Nucl. Instrum. Meth.}, A489:224--229, 2002.

\bibitem{Akimoto:2005rf}
Y.~Akimoto, Y.~Inoue, and M.~Minowa.
\newblock {\em Nucl. Instrum. Meth.}, A557:684--687, 2006.

\bibitem{Moriyama:1998kd}
Shigetaka Moriyama et~al.
\newblock {\em Phys. Lett.}, B434:147, 1998.

\bibitem{Inoue:2002qy}
Yoshizumi Inoue et~al.
\newblock {\em Phys. Lett.}, B536:18--23, 2002.

\bibitem{Inoue:2008zp}
Y.~Inoue et~al.
\newblock {\em Phys. Lett.}, B668:93--97, 2008.

\bibitem{Gattone:1997hf}
A.~O. Gattone et~al.
\newblock {\em Nucl. Phys. Proc. Suppl.}, 70:59--63, 1999.

\bibitem{Bernabei:2001ny}
R.~Bernabei et~al.
\newblock {\em Phys. Lett.}, B515:6--12, 2001.

\bibitem{Morales:2001we}
A.~Morales et~al.
\newblock {\em Astropart. Phys.}, 16:325--332, 2002.

\bibitem{Collaboration:2009ht}
Z.~Ahmed et~al.
\newblock {\em Phys. Rev. Lett.}, 103:141802, 2009.

\bibitem{Lazarus:1992ry}
D.~M. Lazarus et~al.
\newblock {\em Phys. Rev. Lett.}, 69:2333--2336, 1992.

\bibitem{Zioutas:2004hi}
K.~Zioutas et~al.
\newblock {\em Phys. Rev. Lett.}, 94:121301, 2005.

\bibitem{Andriamonje:2007ew}
S.~Andriamonje et~al.
\newblock {\em JCAP}, 0704:010, 2007.

\bibitem{Arik:2008mq}
E.~Arik et~al.
\newblock {\em JCAP}, 0902:008, 2009.

\bibitem{Gondolo:2008dd}
Paolo Gondolo and Georg Raffelt.
\newblock {\em Phys. Rev.}, D79:107301, 2009.

\bibitem{Kaplan:1985dv}
David~B. Kaplan.
\newblock {\em Nucl. Phys.}, B260:215, 1985.

\bibitem{Srednicki:1985xd}
Mark Srednicki.
\newblock {\em Nucl. Phys.}, B260:689, 1985.

\bibitem{Cheng:1995fd}
S.~L. Cheng, C.~Q. Geng, and W.~T. Ni.
\newblock {\em Phys. Rev.}, D52:3132--3135, 1995.

\bibitem{Williams:1971ms}
E.~R. Williams, J.~E. Faller, and H.~A. Hill.
\newblock {\em Phys. Rev. Lett.}, 26:721--724, 1971.

\bibitem{Bartlett:1988yy}
D.~F. Bartlett and S.~Loegl.
\newblock {\em Phys. Rev. Lett.}, 61:2285--2287, 1988.

\bibitem{Ahlers:2007qf}
M.~Ahlers, H.~Gies, J.~Jaeckel, J.~Redondo, and A.~Ringwald.
\newblock {\em Phys. Rev.}, D77:095001, 2008.

\bibitem{Afanasev:2008fv}
A.~Afanasev et~al.
\newblock {\em Phys. Lett.}, B679:317--320, 2009.

\bibitem{Jaeckel:2008fi}
Joerg Jaeckel, Javier Redondo, and Andreas Ringwald.
\newblock {\em Phys. Rev. Lett.}, 101:131801, 2008.

\bibitem{Redondo:2008zf}
Javier Redondo.
\newblock {arXiv:hep-ph/0805.3112}.
\newblock 2008.

\end{thebibliography}
\end{document}